\documentclass[12pt]{article}
\usepackage{graphicx}

\newcommand{\Ref}[1]{(\ref{#1})}
\newcommand{\cft}{$^{\rm 14}$C}
\newcommand{\cth}{$^{\rm 13}$C}
\newcommand{\ctw}{$^{\rm 12}$C}

\newcommand{\INTCAL}{{\sc IntCal}}
\newcommand{\upd}{{\rm d}}

\title{\vspace{-6mm}Is Radioactive Decay Really Exponential?
\vspace{-2mm}
}
\author{Philip J. Aston\thanks{Email: P.Aston@surrey.ac.uk}\\
Department of Mathematics\\
University of Surrey\\
Guildford\\
Surrey GU2 7XH\\
UK}

\begin{document}

\maketitle

\begin{abstract}
Radioactive decay of an unstable isotope is widely believed to be 
exponential. This view is supported by experiments on rapidly decaying 
isotopes but is more difficult to verify for slowly decaying isotopes. 
The decay of \cft~can be calibrated over a period of 12,550 years by 
comparing radiocarbon dates with dates obtained from
dendrochronology. It is well known that this approach shows that
radiocarbon dates of over 3,000 years are in error, which is generally 
attributed to past variation in atmospheric levels of \cft. We note that
predicted atmospheric variation (assuming exponential decay) does not
agree with results from modelling, and that theoretical quantum mechanics
does not predict exact exponential decay. We give mathematical
arguments that non-exponential decay should be expected for slowly 
decaying isotopes and explore the consequences of non-exponential decay. 
We propose an experimental test of this
prediction of non-exponential decay for \cft. If confirmed, a foundation
stone of current dating methods will have been removed, requiring a 
radical reappraisal both of radioisotope dating methods and of 
currently predicted dates obtained using these methods.
\end{abstract}

\section{Introduction}

Radioactive decay is almost universally believed to satisfy the exponential 
decay law over many half lives and in particular cases where it has been 
tested \cite{gopych84,norman88} it has been found to accurately represent the
decay of an unstable isotope. However, as nuclear physics as a discipline 
is little over a century old, these experiments have necessarily been 
restricted to rapidly decaying isotopes. This principle of exponential decay
is then assumed to hold in many other situations where there is no direct 
experimental verification, notably for slowly decaying isotopes which are 
widely used for dating objects ranging from a few hundred years
old, to billions of years old. Newton summarised this situation as follows:
`Although in some instances the exponential law has been experimentally well
verified over many lifetimes, it has certainly not been checked in many
cases in which it is nevertheless assumed to be valid' \cite[p608]{newton66}.


So is there any experimental or theoretical evidence that can be used to 
verify, or disprove, this widely accepted assumption for slowly decaying 
isotopes? 

With regard to experimental evidence, we consider in detail the decay of
\cft~which is the basis of radiocarbon
dating. We propose that non-exponential decay may be a contributing factor 
in the known discrepancy in radiocarbon dates of over 3,000 years.
We also consider the theoretical aspects of radioactive decay as described
by quantum mechanics, and again present arguments for non-exponential decay
for slowly decaying isotopes.
We then consider the consequences if non-exponential decay is assumed 
and conclude with a proposal for an experimental test of non-exponential
decay.

\section{Radiocarbon dating and calibration} 
\label{section2}

The radioisotope \cft~is the basis for radiocarbon dating in which it
is assumed that living organisms have a 
\cft/\ctw~ratio which is the same as that in the atmosphere and that 
this atmospheric ratio has been constant in the past. The interaction between
the organism and the atmosphere stops at death, and the \cft~that remains
in the organism decays. By measuring the \cft/\ctw~ratio directly 
(using Accelerator Mass Spectrometry (AMS)) or the rate of decay of 
\cft~(using radiometric methods), the time since death can be estimated.
Clearly there is another assumption underlying this process, namely that 
the decay of \cft~is exactly exponential for all time. The half-life for 
\cft~was initially found to be $5,568\pm 30$ yr (the Libby half-life) 
but was later changed to $5,730\pm 40$ yr (the Cambridge half life)
\cite{aitken90}. The Libby half-life is still used by convention in 
calculating \cft~ages \cite{aitken90,stuiver77,donahue90}.

Radiocarbon dates can be calibrated using dendrochronology in which the age 
of a sequence of trees of increasing age can be determined very accurately 
by counting the growth rings and cross-dating different samples of
overlapping age \cite{aitken90}. There can be occasional problems with 
missing rings or double rings produced in a single year, but with 
careful analysis and checking of several samples the errors in this 
process are very small (`on the order of 1 yr' \cite{reimer04}). 
The longest continuous tree-ring chronology of $12,593$ years
is based on German oak and pine trees, 
supplemented by pines from Switzerland \cite{friedrich04,schaub08}. 
This approach is unique in providing accurate, 
absolute dates over a period of many thousands of years.

The dates of a tree 
sample obtained using dendrochronology and by radiocarbon dating
are found to be very similar over the initial $3,000$ year 
period, but for older trees there is a significant difference between 
these two dates of up to 15\%. It is assumed that the date obtained from 
dendrochronology is the more accurate one and so these two dating methods 
can be used to generate a calibration curve for converting radiocarbon 
dates to calendar dates. The currently accepted calibration curve is 
\INTCAL09 \cite{reimer09} which is based on dendrochronology for
the first 12,550 years and extends back to 50,000 years using other
methods. We note that radiocarbon dates have units BP (before 
present, where the present is the base year of 1950) or kBP, while 
calibrated dates obtained from the calibration curve have units cal BP 
or cal kBP.

While the use of the calibration curve may have resolved the issue
for the practitioner, who now has a valuable and accurate dating tool,
for the theorist it leaves the obvious question as to what is the cause of
this discrepancy in the radiocarbon dates?

\section{Atmospherics}
\label{atmospherics}

The discrepancy in the calibration curve is generally attributed
to past atmospheric variation in the 
\cft/\ctw~ratio, which contradicts the assumption made in 
the dating process that the ratio was constant in the past.
Four factors are considered to affect this atmospheric ratio, namely
primary cosmic ray flux, strength of the solar electromagnetic field, 
terrestrial magnetic field intensity and the structure of the carbon cycle 
\cite{aitken90,bard98,beck01}.
The combined effect of these factors over the last $12,550$ years is 
not known and since we do not have an independent method of measuring the
atmospheric ratio over many thousands of years, 
it is not possible to correlate the variation in
the atmospheric ratio with the discrepancy in the calibration curve.
However, this is a significant assumption made in the carbon dating process 
which is known not to hold and so is generally considered to be the
sole cause of the discrepancy in the radiocarbon dates.

If the decay of \cft~is assumed to be exponential, then the discrepancy in
the calibration curve can be used to predict the variation $\Delta$\cft~of 
the atmospheric ratio in the past \cite{stuiver98}. Data from 
a stalagmite \cite{beck01} and from marine sediments 
\cite{hughen04} have been used to construct a record of 
$\Delta$\cft~going back to 45 or 50 cal kBP respectively.
These records show much larger variations than are found over the
12.55 cal kBP period that we have considered.
Both studies also used various box models of the global carbon cycle,
which typically did not achieve levels of $\Delta$\cft~as high
as those derived from the data unless extreme values of the parameters were
used. Chiu {\em et al.} \cite{chiu07} analysed these models and
stated that `there is no commonly accepted
explanation for the high atmospheric $\Delta$\cft~values recorded in most
archives'. They also concluded that `the discrepancies between measured
$\Delta$\cft~and modelled $\Delta$\cft~remain unresolved'. They noted that 
one way to reduce the predicted high levels of $\Delta$\cft~would be to 
increase the half life of \cft. A half life of 6,030 years gives results 
that are `entirely consistent with the Beck {\em et al.} (2001) model (A) 
prediction' \cite{chiu07}, which corresponds to an increase in the 
half life of over 5\%. However, recent experimental results have proposed 
that the half life should be {\em decreased} by $2\pm 1$\% \cite{roberts07},
which would of course increase the predicted 
$\Delta$\cft~values even further, and so there is no justification
for the increase in the half life that is required to reduce the high
$\Delta$\cft~values.

While the date range used by these authors is way beyond the range
that we are considering, these studies suggest that variation in the
atmospheric ratio may be insufficient on its own to explain the
discrepancy in the calibration curve. However, allowing for non-exponential
decay as well as atmospheric variation would mean that the predicted
$\Delta$\cft~curve could be reduced to be more in line with the predictions 
obtained from the modelling. So is it possible that some of the discrepancy 
in the radiocarbon dates could be due to non-exponential decay of \cft?

\section{A statistical perspective}
\label{statistics}

Before considering the possibility of non-exponential decay, we first
review the common simple derivation of exponential decay.

Radioactive decay occurs when an unstable atomic nucleus spontaneously
emits energy and matter, often transforming into a new
element in the process. The unstable \cft~nucleus can undergo radioactive 
beta decay
in which an electron and an anti-neutrino are emitted and the \cft~decays
to the stable isotope $^{14}$N. This decay process is assumed to be 
stochastic at the atomic level and hence unpredictable, but the average 
rate of decay for a large number of atoms is often highly predictable. 
Assuming that each atom is equally likely to decay at any time and that the
probability of decay is independent of the age of the atom leads to 
a Poisson process \cite{evans67} which can be described in terms of the 
differential equation
\begin{equation}\label{Alinear}
\dot A=-\alpha A,~~~~A(0)=A_0,
\end{equation}
where $A(t)$ is the \cft/\ctw~ratio at time $t$ with initial value $A_0$
at time $t=0$, which corresponds to the atmospheric ratio,
$\alpha$ is the decay constant which determines the rate of 
decay and $\dot A=\upd A/\upd t$. The solution of equation \Ref{Alinear} is 
$$A(t)=A_0e^{-\alpha t},$$
which gives the expected exponential decay. 

We note that this derivation of exponential decay is based on a simple
statistical argument and
makes no attempt to understand the mechanism that causes the decay. It has
been noted that `the exponential law\ldots is experimentally
verified in some cases to an astonishing degree, much more so than one has
a right to expect considering the rather flimsy basis on which it is
usually derived' \cite[p594]{newton66}. So what is the correct theoretical
basis for understanding radioactive decay?

\section{A quantum mechanical perspective}
\label{quantum-mechanics}

The theory of quantum mechanics describes the process of radioactive 
decay in terms of solutions of the Schr\"odinger equation
\cite{fonda78,merzbacher98}. This theory highlights one of 
the deficiencies of the simple equation \Ref{Alinear},
namely that it does not contain history-dependent terms, as are contained
in the equation for the probability amplitude derived from the theory of 
quantum mechanics \cite{fonda78,peres80}. 

The general solution of the Schr\"odinger equation using first order
perturbation theory does not give the probability of non-decay as
a simple exponential. However, if some
approximations are used, known as the Fermi Golden Rule 
\cite{fonda78,merzbacher98}, then this probability amplitude becomes
exponential. Thus, `the exponential
decay law, for which we have so much empirical support in radioactive decay
processes, is not a rigorous consequence of quantum mechanics, but the
result of somewhat delicate approximations' \cite[p513]{merzbacher98}.

The theory of quantum mechanics predicts variations from pure exponential 
decay at very short times and at very long times \cite{peres80}. However,
these deviations occur in regimes that are currently beyond the reach of
experimental verification \cite{norman88}.

Thus, there is no
theoretical basis from quantum mechanics for exponential decay in
all circumstances. We might therefore ask why the theoretical possibility
of non-exponential decay has not been observed experimentally.

\section{A mathematical perspective}
\label{maths}

One important mathematical property of the differential equation 
\Ref{Alinear} is that it is linear. Ian Stewart observed that `today's 
science shows that nature is relentlessly nonlinear' \cite[p83]{stewart02}. 
This does not mean that there is no room for linear equations
but if the world is essentially nonlinear, then linearity is 
a special case and as such must be justified for a particular problem. 
So could this linear equation in reality be an approximation of
a nonlinear equation? And what difference would it make if this equation
was actually nonlinear?

To address these questions, we note that in the context of radioactive decay, 
there are two quantities that can be used to distinguish four scenarios
namely (i) the rate of decay and (ii) the timescale (relative
to the half life) over which measurements are made. This is illustrated in 
fig.~\ref{table1}. We discuss the terms ``rapid decay'' and 
``slow decay'' in more detail later.

We first consider the short time case. Plotting $A(t)$ on 
a log scale results in a straight line graph if $A(t)$ is exponential.
On a timescale that is small relative to the half life, only a short section 
of the line is observed and this may be approximately straight, 
even if it is not straight over a much longer timescale, so that decay,
when considered over a short time period, appears to be exponential. This 
covers the two possibilities in the bottom row of fig.~\ref{table1}.

\begin{figure}[t]
\begin{center}
\begin{tabular}{r|c|c|}
\cline{2-3}
&&\\[-3mm]
Long time~~ & ~~~~Exponential~~~~ & Non-exponential \\
($\alpha t$ large)~~ & decay & decay? \\
& (Observed) & (Not observed)\\[1mm]
\cline{2-3}
&&\\[-3mm]
Short time~~ & Exponential & Exponential \\
($\alpha t$ small)~~ & decay & decay \\
& (Observed) & (Observed) \\[1mm]
\cline{2-3}
\multicolumn{1}{c}{\vspace{-3mm}} &
\multicolumn{1}{c}{} &
\multicolumn{1}{c}{}\\
\multicolumn{1}{c}{} &
\multicolumn{1}{c}{Rapid decay} &
\multicolumn{1}{c}{Slow decay}\\
\multicolumn{1}{c}{} &
\multicolumn{1}{c}{($\alpha$ large)} &
\multicolumn{1}{c}{($\alpha$ small)}\\
\end{tabular}
\end{center}
\vspace{-5mm}
\caption{The type of decay that is likely in four different cases.}
\label{table1}
\end{figure}

In quantum mechanics, the decay of an unstable state is described by the
Schr\"odinger equation, from which a convolution differential equation can be
derived for a complex transition amplitude $z(t)$ from the initial unstable
state to itself. The probability of finding the system still in the 
initial state at time $t$ is $|z(t)|^2$, which is the quantity that 
is usually assumed to decay exponentially \cite{merzbacher98,peres80}.

The Schr\"odinger equation is linear, which follows from the
Principle of Superposition \cite{merzbacher98}, as is the derived convolution 
differential equation for $z(t)$. However, if we work
with the two real variables $A=|z|^2$ and $\theta=\arg(z)$, then the
differential equations in these two variables are nonlinear. 
Scaling $A$ by $1/A_0$ ensures that it 
satisfies the initial condition $A(0)=1$.
The equation for $A$ can then be expressed in the form
\begin{equation}\label{general-eqn}
\dot A=-\alpha A+g(A,\theta,t),
\end{equation}
where the nonlinear function $g$ involves an integral over past time and
satisfies $g(0,\theta,t)=0$. We are not concerned with the equation for 
the second variable $\theta$. 

We now define rapid decay to be when $\alpha$ is large compared to 
the nonlinear term $g(A,\theta,t)$ and slow decay to be when the two terms 
are of comparable magnitude.
\begin{enumerate}
\item
If decay is rapid, then
the nonlinear term in equation \Ref{general-eqn} will be small compared 
to the linear term resulting in decay that is very close to exponential for
a long time. This covers the top left box of fig.~\ref{table1}.
\item
If decay is slow, then the linear and nonlinear 
terms will be more comparable in magnitude, and the nonlinear 
term will then play an important part in the decay dynamics, resulting
in non-exponential decay. This is the remaining combination in 
fig.~\ref{table1}.
\end{enumerate}

The above discussion contains a number of assumptions which cannot easily be
verified. However, it is undoubtedly true that accurate measurements 
have been made in three cases, as also indicated in fig.~\ref{table1}, 
with all such measurements strongly supporting the assumption of 
exponential decay. For example, the decay of $^{56}$Mn, which has a
half life of 2.5785 h, has been measured experimentally up to 45 half lives
in order to search for non-exponential effects, but none were found
\cite{norman88}. In the fourth case, it has not been possible to 
accurately measure slow decay over long time periods, which
is precisely the case where it is hard to justify the assumption of a linear
equation giving rise to exact exponential decay. 

Thus, it is possible that the slow decay of \cft~could be 
non-exponential, and consequently that this non-exponential decay is a 
contributing factor to the discrepancy in the radiocarbon dates described 
above.

\section{What if?}
\label{whatif}

We have thus far made a case that variation in the atmospheric ratio may be
insufficient on its own to explain the discrepancy in the calibration
curve, and that the possibility of non-exponential decay should also be
considered. We therefore now assume that the discrepancy in the 
calibration curve is {\em a combination of atmospheric variation and 
non-exponential decay}. Of course our big problem,
having made this assumption, is that we do not know how much each of these
two factors contributes to the discrepancy. There are two extremes that
could be considered, namely (i) that the discrepancy is
entirely due to atmospheric variation (and hence that decay is exponential) 
or (ii) that there is no atmospheric variation and the discrepancy is 
entirely due to non-exponential decay. 
Case (i) is usually assumed to apply, and the consequences of this
assumption have been studied, resulting in the $\Delta^{14}$C curve 
which we discussed above. We now make the alternative
extreme assumption described by case (ii), in order to consider the 
consequences of this assumption, whilst recognising that this does not
occur in practice.

\begin{figure}[t] 
\centerline{\includegraphics[width=8cm]{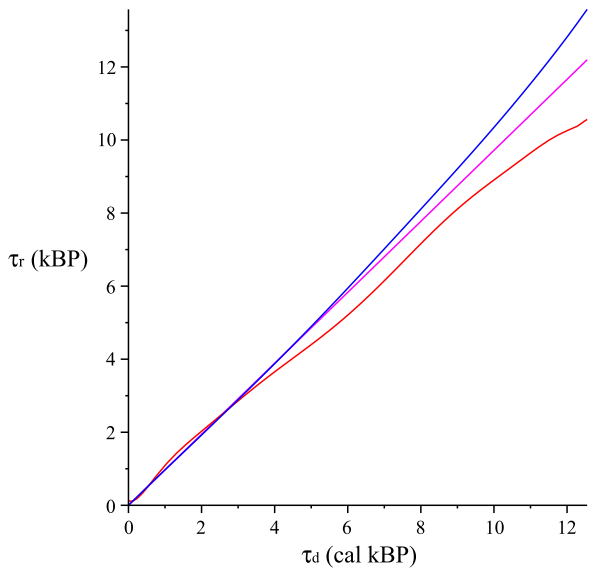}}
\vspace{-2mm}
\caption{The calibration curves $C_\beta$ (red) and $C_{\rm AMS}$
(blue) together with the line $\tau_r=(\alpha_C/\alpha_L)\tau_d$
(magenta).} 
\label{tauamsfig}
\end{figure}

Before continuing, we digress briefly to consider the INTCAL09 dataset
\cite{reimer09} which consists
of an underlying curve with superimposed small oscillations, known
as ``wiggles''. A close correspondence between these wiggles and solar
activity related to sunspots over a period of 400 years has been observed
\cite{bard98,stuiver61}. Thus, we ignore the wiggles and extract the 
underlying trend in the data. Let $\tau_r$ and $\tau_d$ be the dates of 
a tree sample obtained using radiocarbon dating and dendrochronology 
respectively. We fit the data with a polynomial of degree 12
which we denote by $\tau_r=C_\beta(\tau_d)$. This smoothed curve is
shown in fig.\ \ref{tauamsfig}.

There are two methods used for carbon dating, namely radiometric methods 
($\beta$-counting),
which measure the rate of decay, and AMS, which measures the
\cft/\ctw~ratio directly. A consequence of the assumption in case
(ii) is that these two methods will give different predicted dates 
for a given sample. To quantify this difference, let $Y(t)$ be the 
(non-exponential) decay profile of the \cft/\ctw~ratio
which satisfies $Y(0)=A_0$, where $A_0$ is the
(assumed constant) value of the atmospheric ratio. We also define a 
non-dimensional decay function $y(t)=Y(t)/A_0$ which is the proportion of
the original \cft~that remains at time $t$, and satisfies the initial 
condition $y(0)=1$.

With radiometric methods, the sample activity, or rate of decay, is
measured, normalised to take account of isotopic fractionation, and
adjusted to the 1950 level.
This normalised sample activity, denoted by $A_{\rm SN}$, is compared 
with the similarly normalised activity of an oxalic acid standard,
denoted by $A_{\rm ON}$, which is considered to be the same as the activity 
of the atmospheric carbon. The radiocarbon age
$\tau_r$ (BP) of the sample is then given by \cite{stuiver77}
\begin{equation}\label{ager}
\tau_r=-\frac{1}{\alpha_L}\ln\left(\frac{A_{\rm SN}}{A_{\rm ON}}\right),
\end{equation}
where $\alpha_L=\ln 2/5568$ yr$^{-1}$, which involves the Libby half life.

\begin{figure}[t] 
\centerline{\includegraphics[width=8cm]{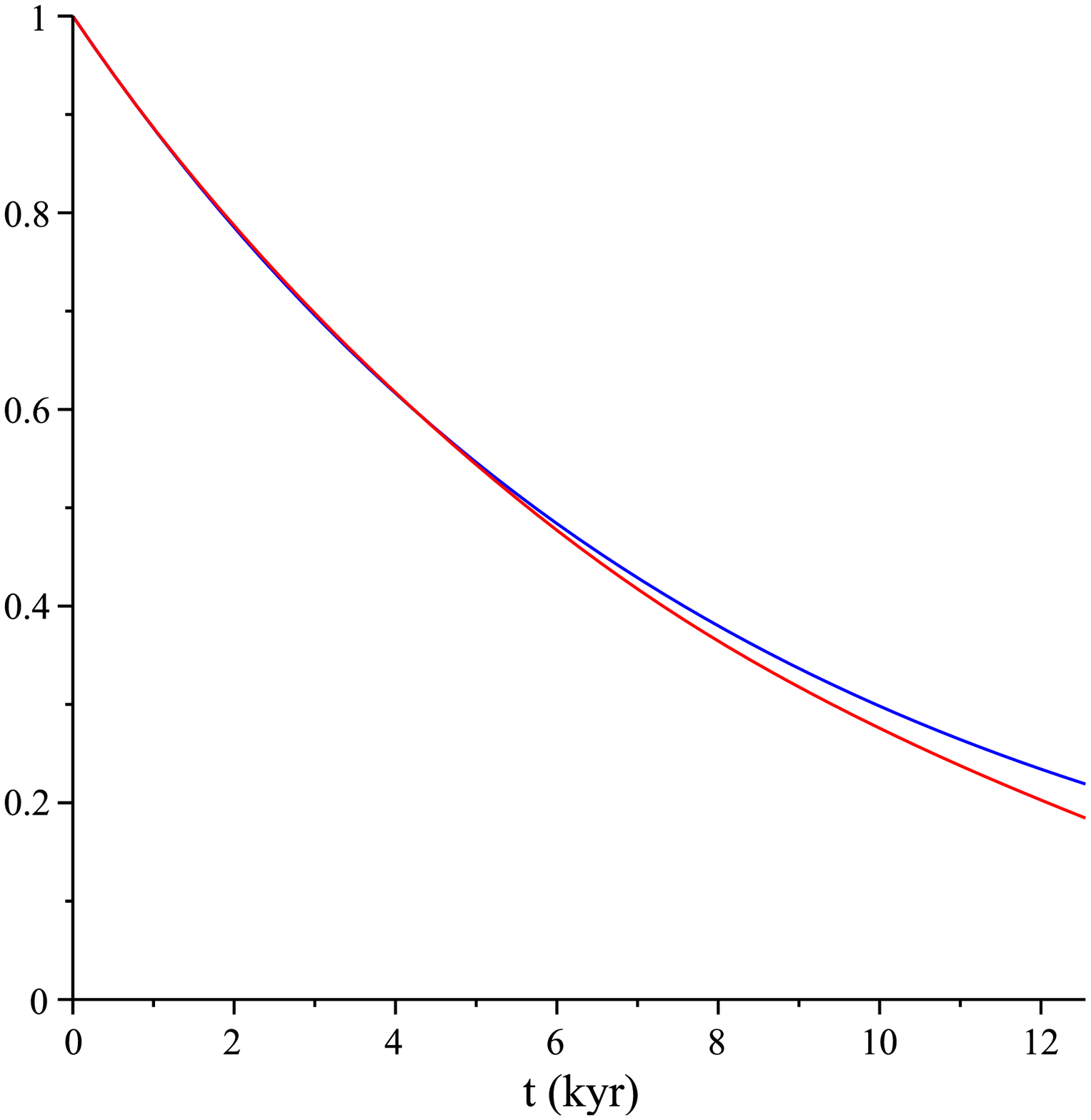}}
\vspace{-2mm}
\caption{$y(t)$ given by \Ref{yt} (red) compared with
the exponential function $e^{-\alpha_C t}$ (blue) with $\alpha_C=
\ln 2/5730~{\rm yr}^{-1}$.}
\label{yplot}
\vspace{3mm}
\centerline{\includegraphics[width=8cm]{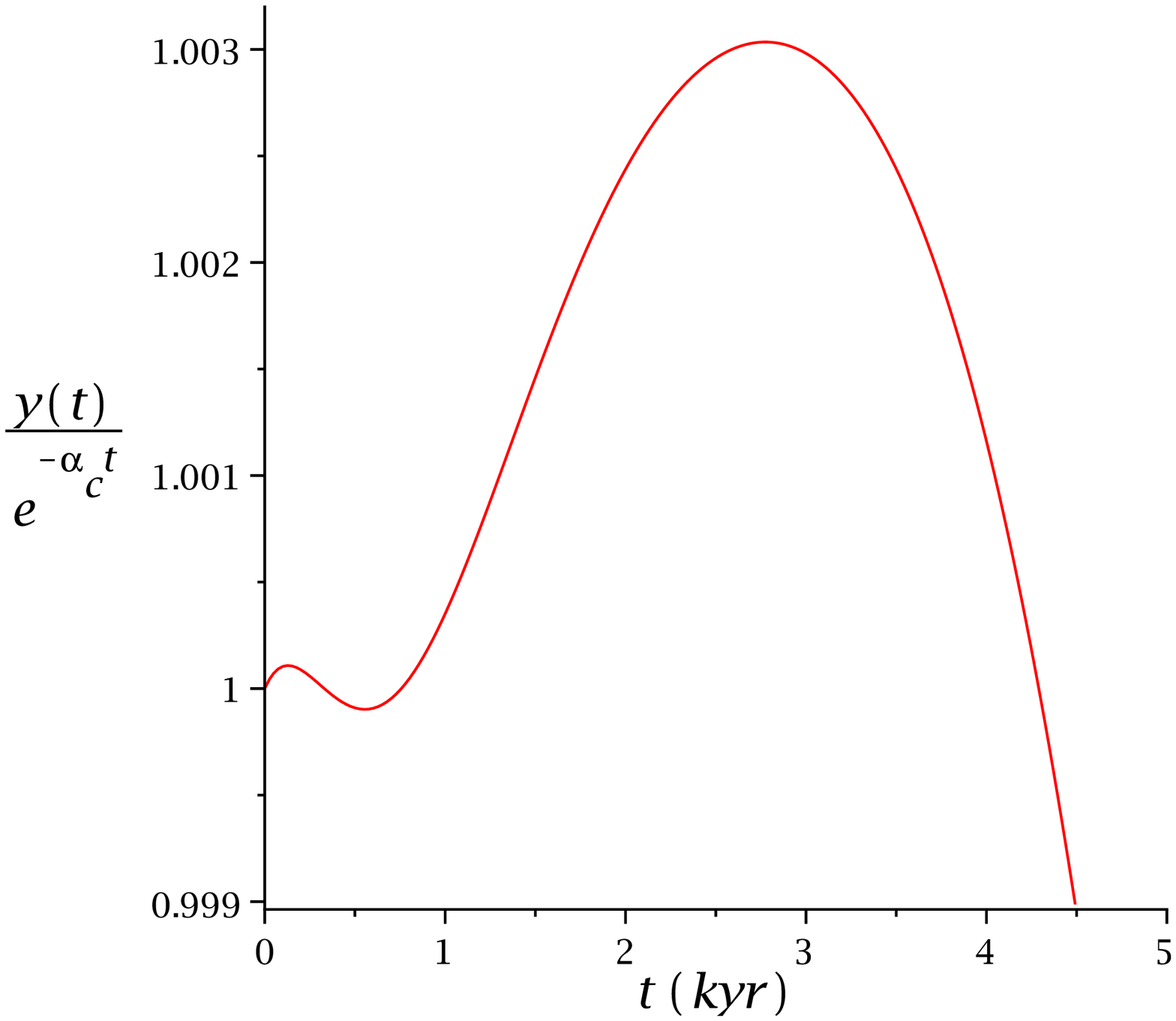}}
\vspace{-2mm} 
\caption{Plot of $y(t)/e^{-\alpha_Ct}$ showing small amplitude 
oscillations.}
\label{oscill} 
\end{figure}

If the decay curve is given by $Y(t)$, then the activity of the sample at
time $t_0$ is $A_{\rm SN}=-\dot Y(t_0)$.
The activity of the standard, assuming that decay is exponential initially,
is $A_{\rm ON}=\alpha_C A_0$,
where $\alpha_C=\ln 2/5730$ yr$^{-1}$, which is the decay constant based on
the more accurate Cambridge half-life. Substituting for $A_{\rm SN}$ and 
$A_{\rm ON}$ in \Ref{ager} then gives
\begin{equation}\label{beta-age}
\tau_r=-\frac{1}{\alpha_L}\ln\left(-\frac{\dot Y(t_0)}{\alpha_C A_0}\right)
=-\frac{1}{\alpha_L}\ln\left(-\frac{\dot y(t_0)}{\alpha_C}\right).
\end{equation}
Rearranging equation \Ref{beta-age} gives
$$\dot y(t_0)=-\alpha_C e^{-\alpha_L\tau_r}.$$
The calibration curve \INTCAL09 in the range of interest of 
0--12.55 cal kBP was derived almost exclusively using radiometric
methods. Using the smoothed calibration curve
we therefore have that $\tau_r=C_\beta(\tau_d)$ where
$\tau_d$ is the true age of the sample obtained by dendrochronology. This
is the same as the time $t_0$, and so we have
$$\dot y(t_0)=-\alpha_C e^{-\alpha_L C_\beta(t_0)}.$$
This equation holds for all values of $t_0$, and so we replace it
with $t$. Integrating and using the initial condition $y(0)=1$ then gives
\begin{equation}\label{yt}
y(t)=1-\alpha_C \int_0^t e^{-\alpha_L C_\beta(s)}~\upd s.
\end{equation}
This function is compared with the corresponding exponential decay curve
$e^{-\alpha_C t}$ in fig.~\ref{yplot}, 
from which it can be observed that there is very good agreement between the
two curves initially with more of a difference developing after
approximately 6,000 years.

The ratio of the function $y(t)$ and the exponential 
$e^{-\alpha_Ct}$ shows small amplitude oscillations initially (see
fig.~\ref{oscill}). These arise due to the fact that the calibration 
curve initially
oscillates also (see fig.~\ref{tauamsfig}). In a simple tunnelling problem,
the survival probability was found to be non-exponential, and also showed
oscillations \cite{dicus02}, although the amplitude was much larger than
those shown in fig.~\ref{oscill}.

We note that if it is assumed that decay is exponential,
then substituting $y(t_0)=e^{-\alpha_C t_0}$ into \Ref{beta-age} and
replacing $t_0$ with the dendrochronologically determined age $\tau_d$
gives 
$$\tau_r=\frac{\alpha_C}{\alpha_L}\tau_d=0.9717\tau_d,$$ 
which is slightly lower than the line $\tau_r=\tau_d$. The reason for this 
is the historic use of $\alpha_L$ rather than $\alpha_C$ in \Ref{ager}.

Using AMS, both the \cft/\cth~and \cth/\ctw~ratios of the sample and a
standard are measured. The \cft/\cth~ratio is normalised to take account
of isotopic fractionation, which adjusts the \cth/\ctw~ratio of the
sample to be the same as that of the standard, and an adjustment to the 1950
level is made. The normalised \cft/\cth~ratios 
of the sample and the standard are denoted by $(14/13)_{\rm SN}$ and 
$(14/13)_{\rm ON}$ respectively and the radiocarbon age $\tau_r$ (BP) is 
then given by \cite{donahue90}
\begin{equation}\label{ams-eqn}
\tau_r=-\frac{1}{\alpha_L}\ln\left(\frac{(14/13)_{\rm SN}}{(14/13)_{\rm ON}}
\right),
\end{equation}
where, by convention and for consistency with \Ref{ager}, the Libby half life
is again used for the decay constant. Again assuming that the true decay 
curve for the \cft/\ctw~ratio is given by $Y(t)$, then we have that 
$(14/13)_{\rm SN}/(14/13)_{\rm ON}=Y(t_0)/A_0$, where $t_0$ 
is the true age, since the \cth/\ctw~ratios of the sample and the standard
are the same. Thus, equation \Ref{ams-eqn} becomes
\begin{equation}\label{AMS-age}
\tau_r=-\frac{1}{\alpha_L}\ln\left(\frac{Y(t_0)}{A_0}\right)
=-\frac{1}{\alpha_L}\ln y(t_0).
\end{equation}
Now $t_0$ is the true age of the sample, which we denote by $t_0=\tau_d$,
and so from equation \Ref{AMS-age} we now get a new calibration curve
based on AMS measurements given by
\begin{equation}\label{camseqn}
\tau_r=C_{\rm AMS}(\tau_d)=-\frac{1}{\alpha_L}\ln y(\tau_d),
\end{equation}
where the function $y$ is given by \Ref{yt}. This curve is shown in 
fig.~\ref{tauamsfig},
together with the line $\tau_r=(\alpha_C/\alpha_L)\tau_d$ which is 
associated with exponential decay. From this, we see that if
assumption (ii) holds then AMS measurements should give consistently 
older radiocarbon dates than those obtained using radiometric methods for 
$\tau_d>3$ cal kBP. 

While the predicted $\Delta^{14}$C curve which arises from 
case (i) cannot easily be verified experimentally,
our prediction from case (ii), that radiocarbon
dates obtained using the two different methods should be significantly
different for older samples, is very easy to check using results
of the Fourth International Radiocarbon Intercomparison (FIRI)
\cite{scott03}.
These results show no significant difference in the predicted ages of a
range of standard samples, including a sample of humic acid (sample E)
dated at around 11.8 kBP, which is just beyond the range that we are
considering. The AMS mean age for this sample was $54\pm 53$ \cft~yr higher 
than the corresponding age obtained using Gas Proportional Counting, a
radiometric method, which is nowhere near the approximately 3,000
\cft~yr difference that our results predict for this sample.

These results would appear to invalidate the assumptions of case (ii) and
support the conventional (but not verified) assumptions of case (i). 
However, before dismissing case (ii), we make two proposals 
that might explain why this predicted difference in dates is not observed.

Firstly, case (ii) is the extreme case. In the more likely scenario in 
which the discrepancy in the calibration curve is due to a combination of 
atmospheric variation and non-exponential decay the predicted difference 
between $C_{\rm AMS}$ and $C_\beta$ would be smaller.

Secondly, we note that radiometric methods were the first to be used for 
carbon dating. In the late 1970's,
AMS techniques were developed to measure the \cft/\ctw~ratio directly
\cite{fifield99}. However, `absolute AMS isotopic measurements are extremely 
difficult' \cite{nishiizumi07} and so these methods were
calibrated using international standards and modern samples of known age.
For older samples, and given a strong belief in exponential decay, the
results were also compared with the predicted radiocarbon ages obtained 
by radiometric methods.
Indeed, the first aim of the FIRI exercise was `demonstration
of the comparability of routine analyses carried out by both AMS and
radiometric laboratories' \cite{scott03}.

In practice, AMS measurements are the product of complex experimental
procedures \cite{fifield99}. Results are affected by a number of factors,
such as contamination during chemical processing 
\cite{donahue90,brown97,mueller02}, sample size \cite{brown97,brown01} and 
pretreatment of samples by acid hydrolysis \cite{brown01,scott03}. 
A small change in the AMS measurement as a result of any of 
these factors would be sufficient to generate a significant difference in 
the predicted radiocarbon age for older samples.

We therefore conclude that as long as AMS measurements are calibrated to give 
agreement with results from radiometric methods, our predicted
difference in radiocarbon dates between the two methods will never be 
observed.

\section{Conclusions}
\label{conclusion}

The question posed in the title of this paper as to whether radioactive 
decay is really exponential (in all circumstances) is of fundamental 
importance, both theoretically and practically. We have presented 
experimental evidence, from the results of radiocarbon dating and modelling,
which suggests that non-exponential decay may be a significant factor in 
the discrepancy in radiocarbon dates. We have also noted that quantum
mechanics does not predict exact exponential decay, and have presented
mathematical arguments that non-exponential decay should be expected for
slowly decaying isotopes, but not for rapidly decaying isotopes.
Thus, we believe that there is strong evidence, both experimental and
theoretical, that radioactive decay of slowly decaying isotopes is {\em
not} exactly exponential. 

We have also shown that one significant consequence of non-exponential 
decay is that the two methods used for
radiocarbon dating, namely radiometric and AMS, should give different
results for older samples. This effect is not seen in practice, but we 
suggest that this could be due to AMS being calibrated to give agreement 
with results obtained using radiometric methods. 

Our results in the previous section provide a way to experimentally 
test our prediction that the discrepancy in the calibration
curve is due to a combination of atmospheric variation and non-exponential
decay. To do this, a careful calibration of AMS measurements is required 
to ensure that correct absolute measurements are being made, particularly 
in the range 3--12.55 cal kBP. Dating of tree samples in this range should 
then be performed. If these radiocarbon dates are significantly higher than
those in the \INTCAL09 dataset, then it would confirm our prediction.
Conversely, if good agreement is obtained with the \INTCAL09 data, then it
would disprove our prediction.

The question we are addressing of whether radioactive decay is
exponential over long time periods is certainly of much theoretical
interest. However, the stakes are high in more practical realms, since
many dating methods over long time periods rely on the
assumption of exponential decay of a slowly decaying isotope
\cite{aitken90}.
If the decay of \cft~is indeed non-exponential, then no other slowly
decaying isotope can be assumed to decay exponentially either, which would
remove a foundation stone of modern dating methods. If confirmed, this will
require a radical reappraisal both of our approach to radioisotope dating
methods and of the many currently accepted dates which have been obtained 
using these methods.
\vspace{-2mm}

\subsection*{Acknowledgements}
I am grateful to Professor Ron Johnson for many helpful discussions and 
comments and his continued interest in this work, to Professors Bill 
Gelletly, Jeff Tostevin 
and Phil Walker for commenting on an early draft and to Professor Michael 
Kearney for some helpful comments.



\end{document}